# Approche conceptuelle par un processus d'annotation pour la représentation et la valorisation de contenus informationnels en intelligence économique (IE).

Sahbi SIDHOM

Maître de Conférences & Chercheur équipe SITE du LORIA

LORIA - Université de Nancy2, BP. 239, 54506 Vandoeuvre - France
Sahbi.Sidhom@loria.fr

## RÉSUMÉ

À l'ère de la société de l'information, l'impact des systèmes d'information sur l'économie du matériel et de l'immatériel est indéniablement perceptible. Vis-à-vis des ressources informationnelles d'une organisation, l'annotation s'implique pour enrichir un contenu informationnel, pour retracer l'activité intellectuelle sur un document et pour mettre de la valeur ajoutée sur l'information au profit de la résolution d'un problème décisionnel dans le contexte de l'intelligence économique.

Notre contribution se distingue par la représentation d'un *processus d'annotation* ainsi que les concepts qui y sont inhérents pour aider le décideur dans sa prise de décision. Il s'agit de la disposition d'une information pertinente qui a été annotée.

Une telle information dans le système est facilitée en prenant en compte la diversité des ressources et pertinemment celles qui sont annotées de manière formelle ou informelle par les acteurs en IE. Une orientation capitale dans ce travail consiste à intégrer dans la prise de décision l'activité de l'annotateur, l'agent logiciel (ou les mécanismes de raisonnement) et la valorisation de ressources informationnelles.

### MOTS-CLÉS :

Processus d'annotation, inférence logique, gestion de connaissances, problème décisionnel, système d'information (SI), intelligence économique (IE).

## ABSTRACT

In the era of the information society, the impact of the information systems on the economy of material and immaterial is certainly perceptible. With regards to the information resources of an organization, the annotation involved to enrich informational content, to track the intellectual activities on a document and to set the added value on information for the benefit of solving a decision-making problem in the context of economic intelligence.

Our contribution is distinguished by the representation of an annotation process and its inherent concepts to lead the decisionmaker to an anticipated decision: the provision of relevant and annotated information.

Such information in the system is made easy by taking into account the diversity of resources and those that are well annotated so formally and informally by the EI actors. A capital research framework consist of integrating in the decision-making process the annotator activity, the software agent (or the reasoning mechanisms) and the information resources enhancement.

______________

______________

# Approche conceptuelle par un processus d'annotation pour la représentation et la valorisation de contenus informationnels en intelligence économique (IE).

## 1. INTRODUCTION

Une définition usuelle d'un SI ressemblait à l'ensemble *des informations formalisables* circulant dans l'entreprise qui sont caractérisées par des liens de dépendance*, des procédures* et *des moyens nécessaires* pour les définir, les rechercher, les formaliser, les conserver et les distribuer. Mais cette définition n'indique ni « à quoi ? » sert le SI, ni « comment ? » il est construit et ni « pourquoi ? » pour quantifier sa « dynamique ». Pour y remédier, il fallait bien distinguer des facettes dans un SI : *(i)* la première, elle est orientée vers les moyens (ie. le système informatique), *(ii)* la seconde, vers les besoins et les usages, auxquels la réflexion sur le SI donne désormais des performances dans la vie de l'entreprise et *(iii)* la dernière, vers la valeur ajoutée en termes de capitalisation de connaissances, savoir et savoir-faire de l'entreprise.

Dans une approche qui étaye notre point de vue, un SI peut être une construction dynamique (Turki, 2005) formée de :

- – informations,
- – traitements,
- – règles d'organisation et
- – ressources dynamiques (humaines et technologiques).

L'ensemble des **informations** intègre des représentations partielles ou des faits (implicites et explicites) qui intéressent l'entreprise. Les **traitements** constituent des procédés d'acquisitions, de mémorisation, de transformation, de recherche, de présentation et de communication des informations. Les **règles d'organisation** régissent l'exécution des traitements informationnels. Les **ressources** dynamiques (humaines et technologiques) sont ce qui est requis pour le fonctionnement du SI.

Dans le contexte de l'intelligence économique (IE), nous nous intéressons au domaine des SI comme supports aux activités humaines dites « métiers ». Cette catégorie regroupe les SI des institutions, des entreprises, des administrations, etc. (ie. les organisations en général). Intrinsèquement, les SI qui ont une influence directe sur les performances, la compétitivité et la cohésion (ie. les tâches collaboratives) de l'organisation dans un environnement économique en croissante concurrence, tout en prenant en compte la complexité des tâches, l'accès aux technologies avancées et l'adaptation de la stratégie compétitive.

L'objectif dans cette recherche est de proposer une approche conceptuelle pour intégrer dans un SI les actions de l'annotateur, qui est un acteur en IE (veilleur, analyste, décideur, etc.) dans son environnement organisationnel. Il est question d'augmenter la réactivité d'un SI face aux informations demandées par l'acteur en IE : des ressources informationnelles (formelles ou informelles) de l'entreprise, des connaissances partagées par les acteurs, des expertises, des lignes d'action ou des procédures pré-décisionnelles, etc. (Cauvet et Rosenthal, 2001).

Le tout se résume par des ressources déployées au profit de l'acteur en IE à retracer son activité sur les informations, à capitaliser ses connaissances aussi bien son savoir-faire dans une approche collaborative avec d'autres acteurs. Pour l'essentiel, l'IE est un cycle d'informations dont la finalité est « *la production de renseignements stratégiques et tactiques à "haute valeur" ajoutée* » (Besson et Possin, 2006), facilitant ainsi « *la maîtrise et la protection de l'information stratégique pertinente par tout acteur économique* » (Juillet, 2007).

L'objet d'étude des SI confirme donc une nouvelle relation dont les principaux enjeux sont liés à l'IE, par les dimensions : veille, capitalisation, protection du patrimoine et influence. Les enjeux en question permettent d'assurer la réactivité de l'entreprise face à son environnement (Le Moigne, 1990) et sa communication avec ses acteurs et ses marchés (ie. biens matériels et immatériels) tout en mesurant les risques, les menaces et son potentiel face à ses concurrents.

Dans ce travail, notre contribution est orientée vers la valorisation de contenus informationnels (ie. formels et informels) avec l'apport coopératif (ie. entre les acteurs en IE) et collaboratif (ie. avec les outils et les applications dédiés à l'IE) dans la production de l'information (Kislin, 2007) et de la valeur ajoutée par le biais d'un processus d'annotation. Il est question d'établir une meilleure connexité entre les contenus informationnels valorisés et les acteurs, les décideurs en IE face à leur problème décisionnel.

## 2. VALORISATION DE CONTENUS INFORMATIONNELS

Dans le prolongement des travaux de recherche sur la valorisation de contenus informationnels, il est essentiel de concourir vers une approche permettant d'accéder aux éléments hétérogènes d'un document, d'analyser leur contenu et d'en extraire les informations pertinentes par l'application de divers processus : indexation, extraction de connaissances, formalisation de concepts, intégration d'annotation, structuration d'objet informationnel, etc. (Bachimont, 2003), (Sidhom et David, 2006). Par conséquent, la disposition de telles informations dans un SI est capitale dans le cycle décisionnel (ie. procédures de décision) d'une organisation face aux problèmes (ie. informationnels, stratégiques et décisionnels).

Particulièrement, dans des travaux sur le filtrage et l'extraction automatique de concepts (ie. indicateurs, termes, index), certaines propositions intéressent de près les acteurs de l'ingénierie documentaire, comme ceux de la gestion de connaissances ou de l'IE. En exemple, dans le contexte de veille, un acteur en IE (ie. veilleur ou analyste) est amené à : –développer des *indicateurs* sur son problème décisionnel puis à les traduire en *termes* et *équations de recherche* pour effectuer la recherche d'information ; –filtrer les *documents* qui répondent au mieux à ses requêtes ; –annoter les *informations pertinentes* de sa recherche par rapport aux besoins informationnels exprimés. Dans ces transitions, cet acteur progresse avec des objets construits dans son activité : indicateurs, termes et équations de recherche, documents, informations pertinentes (ie. jugées pertinentes par l'acteur) et leurs annotations. Cet enchaînement ne peut évoluer que vers des informations à valeur ajoutée dans un SI : du document présent dans la base du SI (ie. l'information primaire), à l'information pertinente (ie. information sélectionnée par le système et approuvée par l'acteur) puis son annotation (ie. son enrichissement). Ainsi, les outils d'analyse, d'indexation et d'annotation seront incontournables sur le marché économique pour disposer d'une telle complétude entre les catégories de l'information afin d'élaborer des stratégies décisionnelles adéquates à l'environnement de l'entreprise et ses besoins.

Dans notre approche, nous émettons un regard critique sur le concept "annotation", qui de notre point de vue implique deux acceptions : c'est à la fois **l'objet** qui matérialise le *contenu de l'annotation* et **l'action** qui identifie *l'activité de l'annotateur* pour la mise en valeur d'un contenu informationnel en intégrant de nouveaux éléments informationnels, relationnels (entre les contenus) ou interprétatifs sur le document.

Dans cette considération, premièrement, l'annotateur est le producteur de l'objet d'annotation et que son activité s'intègre "globalement" dans un *processus interprétatif* sur le contenu informationnel. Il s'agit d'un processus interprétatif dès qu'un annotateur relie "*toute information simple ou complexe à une autre avec une singulière signification*" au moment de sa matérialisation. Le processus pourra faire appel aux préceptes d'analyse, d'indexation ou de concrétisation des percepts (ie. les objets de la perception) par une lecture synthétique et explicative sur le contenu documentaire (Bachimont, 2004). Deuxièmement, l'annotation est un objet (écrit, graphique, sonore ou multimédia) qui est rattaché au document pour préserver l'*aspect complémentaire* entre la nature des informations sources (ie. le document) et celles introduites à valeur ajoutée (ie. l'annotation).

En vertu de cette présentation sur les aspects de la valorisation informationnelle (ie. objet, action et processus) d'un contenu, nous appuyons, dans ce qui suit, notre approche sur l'**objet** par l'importance d'des visées sémantiques dans l'annotation et l'implication du concept de la valeur ajoutée pour l'aide à la décision en IE.

## 2.1. Accès à l'information : connaissances explicites

Les informations fournies par un système de recherche d'information (SRI) amènent l'acteur en IE à les valider (ie. pertinence de certaines informations sur l'ensemble), à annoter ses résultats en adéquation avec les besoins informationnels exprimés. Dans cette interaction (acteur, système et documents), la fiabilité des résultats est conjointement liée à :

–   la structure, la typologie des documents et leur gestion par le système pour devenir accessibles selon le mode de lecture de l'utilisateur (ie. ses préférences) ;

–   les thèmes et les concepts appropriés par le document selon la culture de l'utilisateur et ses connaissances du sujet (culturel, scientifique, technique, etc.) ;

–   le contexte de la recherche d'information (RI) défini pour le système (ie. la formulation des requêtes) par analogie avec le contexte de l'utilisateur (ie. en IE, de l'explicitation d'un problème décisionnel à sa traduction en un problème de recherche d'information) ;

C'est ainsi qu'en IE, l'acteur-annotateur en phase de RI peut être amené à accomplir des interprétations collaboratives et coordonnées avec d'autres afin de comprendre le problème décisionnel posé, selon l'explicitation du décideur, pour le traduire en un problème de RI et de retourner au décideur des informations utiles sur le problème.

Dans un tel contexte, l'annotation permet d'introduire cet aspect interprétatif sur l'information selon les connaissances, le savoir et le savoir-faire de cet acteur-annotateur. Justement, nous avons trouvé plusieurs définitions attribuées à l'annotation et les plus significatives, de notre point de vue, accordent un aspect évolutif du concept, à savoir :

–   " *bref commentaire ou explication sur un document (ou son contenu), de même une très brève description habituellement ajoutée en note après la référence bibliographique du document.* " (GTD, 2007) ;

–   " *une annotation est une information graphique ou textuelle attachée à un document et le plus souvent placée dans ce document.* " (Desmontils et alii, 2004) ;

–   " *tout objet qui est relié à un autre par un certain rapport (relation, corrélation ou lien à définir)* ", *(ie. Any object that is associated with another object by some relationship).* (W3C-Annotation, 2004) ;

Ainsi, l'annotation se caractérise par différentes dimensions relatives à l'objet « annotation ». Les dimensions de l'objet doivent décliner ses propriétés, à savoir : sa structure, ses fonctions et son rôle dans la communication entre les acteurs en IE ou entre l'acteur et le SI. Dans ce contexte, trois éléments essentiels sont à prendre en considération, à savoir :

–   le producteur de l'annotation : *le profil de l'annotateur,*
–   le contenu documentaire concerné par l'annotation : *l'information source,*
–   l'objet d'annotation introduit sur le contenu : *les (informations à) valeurs ajoutées.*

Par acception, un document est une trace de l'activité humaine, c'est une considération que nous la retenons d'un point de vue de l'effort intellectuel humain pour représenter des faits, des connaissances et du savoir-faire dans son environnement socio-culturel, économique, etc. A cet égard, l'effort interprétatif humain assigné à un contenu informationnel cadre bien cette acception « documentique ».

Or, une base de données, quelque soit les caractéristiques (ie. données hétérogènes, multi-formes, multi-sources, etc.), permet de mettre ses données à la disposition de ses utilisateurs pour une consultation, une saisie ou une mise à jour tout en s'assurant des droits d'accès et des privilèges accordés. Ses données contenues ont été *structurées, formatées, typées, contrôlées et indexées* pour une mise à disposition et à l'exploitation : c'est ce qui caractérise l'information formalisée.

Certes autour du concept d'indexation (ie. données, informations ou documents), des propos retenus par Jacques Maniez (Maniez, 2002-2007) qui a repris la définition du dictionnaire "Le Robert" en la commentant par :

> « *Indexer dit le Robert c'est attribuer à un document une marque distinctive renseignant sur son contenu et permettant de le retrouver* ».

> Puis, « *Les termes indice, index et indexation appartiennent à la même famille et les différences subtiles qui les relient entre eux méritent d'être explicitées* ».

> Et il complète son analyse par : « *Indexer au sens documentaire du terme, c'est accoler sur le document un indice, au sens large de repère, autrement dit une étiquette permettant de repérer son sujet matière...* ».

Concernant les différentes stratégies utilisées par les indexeurs, leurs traitements sur les documents ne peuvent être réalisés en dehors du **territoire expérimental** : la méthodologie et le contexte (El-Hachani, 2005) : c'est ce qui se projette dans le processus d'annotation. Ce dernier se relaye avec l'indexation avec plus de *« liberté »* accordée à l'annotateur (ie. **la méthodologie**) et de *« souplesse » accordée* au contenu de l'annotation et de sa fiabilité (ie. sa pertinence dans **le contexte**), pour faciliter la navigation dans les contenus en leur attribuant de la valeur ajoutée par le processus d'annotation.

L'objectif étant clair autant pour l'objet d'indexation que celui d'annotation, c'est de représenter des connaissances, le plus explicitement possibles, sur le document et de permettre une recherche d'information pertinente. En pratique, la formalisation des concepts permet de faciliter l'exploitation (manuelle ou automatique) de documents, des informations contenues et des connaissances associées.

## 2.2. Valorisation par l'annotation : valeurs ajoutées

Différentes études traitant la gestion documentaire et informationnelle mettent en avant des principes de capitalisation des connaissances afin d'en faciliter l'accès aux ressources. Seulement cette mise à disposition nécessite, en amont, la gestion de contenus, l'analyse de l'information et le management des connaissances et, en aval, des critères sur le contexte d'utilisation et sur les besoins liés à l'information. Ce qui engage la prise en compte des paramètres suivants :

- **l'exhaustivité :** étendre l'analyse du document du sujet principal aux sujets connexes ;

- **l'indexation orientée utilisateur :** élargir l'analyse du contenu aux éléments implicites du document qui peuvent être utiles à l'acteur en IE ;

- **l'information nouvelle :** inclure, lors de l'analyse du document, des choix de concepts complémentaires au concept principal, des réponses à des problématiques posées dans le sujet (théories, hypothèses, etc.), des informations intra- ou inter- linguistiques par rapport à l'information initiale, etc. ;

- **la base d'intérêt :** tenir compte, sur le contenu d'un document, les idées adoptées par l'auteur et les points de vue transmis par les annotateurs : à déterminer le profil de chaque acteur (auteur, annotateur, décideur, etc.) ;

- **la nature de l'information :** distinguer la catégorie **formelle** ou **informelle** de l'information. A posteriori, l'information explicitée dans un document ne reflète pas directement l'intérêt recherché par un veilleur pour répondre à un besoin informationnel. Certes, les concepts d'équivalence linguistique entre un document et un veilleur ne sont pas formellement définis. Dans ce cas de figure, une annotation peut enrichir ce document, projeter de nouvelles idées (ie. intérêt, objectivité ou subjectivité), rapprocher ou éloigner les points de vues des annotateurs et auteurs, définir des concepts d'équivalence, etc. Ces aspects intéressent de près les décideurs sur des évènements non communiqués "officiellement" ou pour identifier l'avis de nouveaux acteurs et collaborateurs.

De ce point de vue, une annotation s'associe *« théoriquement »* à des informations hétérogènes pour rajouter *« pratiquement »* de nouvelles informations interprétées : en terme des valeurs ajoutées (*cf.* Fig. 1.). Elle est produite dans l'objectif d'être propre à son annotateur ou d'être partagée. Sa lecture est perçue de manière objective ou subjective, car la perception des concepts d'annotation est liée à la prise en compte du facteur humain : le profil de l'acteur en IE (Bouaka, 2004), (Robert et David, 2006).

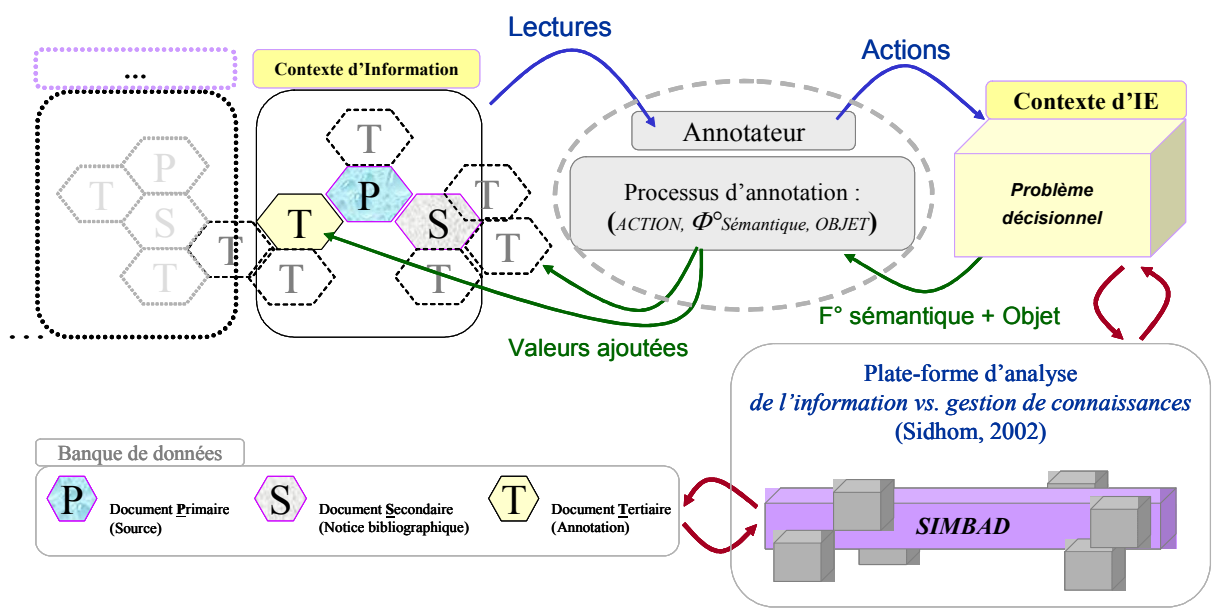

***Figure 1. :** La chaîne de production des annotations : concepts et valeurs ajoutées.*

Les éléments à valeur ajoutée vont faciliter le transit de l'information vers son environnement d'application : dans −un contexte spécifique avec −une disposition d'informations pertinentes pour −aider à la résolution d'un problème décisionnel. Ce raisonnement nécessite de rendre fonctionnelle l'annotation aux acteurs en IE comme aux agents logiciels d'un SI pour le traitement (Sidhom et alii, 2005).

Ainsi, tout document peut disposer des fonctionnalités de l'annotation pour couvrir des termes, des concepts, des connaissances et pour s'allier avec des éléments terminologiques, des liens vers d'autres média, des objets homogènes ou hétérogènes, etc. afin de constituer la valeur ajoutée : c'est la conséquence logique de l'évolution du concept et du processus (David et Sidhom, 2005) avec l'avènement de l'information numérique et de ses pratiques.

## 3. PROCESSUS D'ANNOTATION

Par emprunt à la logique aristotélicienne, le syllogisme est un raisonnement logique à deux propositions (ie. les prémisses) conduisant à une conclusion qu'Aristote a été le premier à formaliser. Par l'emploi de cette logique, les deux prémisses de l'annotation s'identifient à l'action (ie. action/activité d'annoter) et à l'objet (ie. contenu de l'annotation) sur le document. Les deux prémisses sont des propositions supposées vraies qui vont permettre de vérifier la véracité formelle de la conclusion. Pour nous, la conclusion n'est autre que **l'information à valeur ajoutée** sur le document qui expose le rapport entre l'**action** et son **objet**.

Dans l'activité de l'annotateur, l'enrichissement sur le contenu d'un document se matérialise par des informations **explicites** ou **implicites**. Ces informations traduisent des interprétations propres à l'annotateur selon son profil, son environnement organisationnel, ses objectifs, etc. Ces éléments avec le contenu de l'annotation sont intégrés dans l'objet d'annotation.

Dans le processus d'annotation, l'appel à une fonction sémantique par l'action de l'annotateur permet de déterminer l'objet annotation. Cette fonction va servir à clarifier le mode de lecture, d'analyse, d'interprétation, d'indexation, etc. du contenu de cet objet. On cite quelques propriétés de la fonction sémantique, à savoir :

- partager : lecture, écriture ou mise à jour sur l'objet ;
- inclure : marqueur d'analyse, étiquette ou tag pour distinguer une information, une connaissance, etc. ;
- filtrer : repérage, marquage, extraction, etc. sur un contenu ;
- indexer : termes, concepts, thèmes, attribut ou (liste de) valeur(s), etc. ;
- faciliter : structuration, aiguillage, plan, catalogage, etc. pour la lecture ou la compréhension d'un document, d'une information, d'un objet, etc. ;
- attacher : identifiants sémantique, symbolique ou alphabétique sur un contenu, un lien vers un autre contenu (ie. lien hyper- texte/média), etc. ;
- etc.

Egalement, quelques traits du contexte d'utilisation par :

- construction d'une représentation externe au contenu d'un document ;
- insertion d'un élément d'évaluation sur le document (Kelly et Teevan, 2006) : témoignage, apport, constat, démonstration, réfutation, raisonnement, etc. ;
- traçabilité sur un contenu ou sur le format de document : synopsis, point de vue de l'annotateur, modalité d'exploitation, etc. ;
- proposition d'une métrique : pour calculer une fréquence, une corrélation, etc. ;
- etc.

Ainsi, l'implication de la fonction sémantique permet la prise en compte des concepts "*intensionnels*" (ie. propriétés du concept) de l'annotateur et rendre possible les aspects "*extensionnels*" (ie. interprétations ou réalisations possibles) de l'objet. Dans cette approche, un mode de raisonnement est donc nécessaire pour adapter l'information à son environnement par l'intermédiation de l'annotateur. L'élément informationnel annoté peut prendre deux états différents, dans deux environnements en même temps et que la présence d'un observateur (ie. en exemple, l'acteur en IE) peut modifier la réalité observée (Brassac, 2007). Ce propos retenu de la Sémantique Générale (Vanderveken, 1988) retrouve bien son application dans le contexte de l'IE, comme pour informer et désinformer en utilisant la même information (ou le même concept) associé(e) à un procédé de manipulation de l'information.

Cet aspect nous conduit à présenter à la formalisation de l'objet annotation et son intégration dans l'environnement de l'IE, tout en explicitant les principes de notre modélisation.

### 3.1. Formalisation de l'objet « annotation »

Au niveau théorique pour notre modélisation, *l'annotation a été structurée pour un processus de recherche d'information*. Il s'agit de prendre en considération les valeurs ajoutées avec les contenus informationnels d'une organisation (ie. valoriser l'opinion des acteurs en IE sur la pertinence ou non de l'information) dans le contexte d'un problème décisionnel : la fonction sémantique a été définie dans le processus pour *"relier par un certain rapport"* le document aux objets de l'annotation.

Au niveau pratique, nous avons observé que l'intégration d'une annotation assimile trois dimensions, à savoir :

- **la dimension explicitation :** une annotation ne suffit pas à elle-même, elle est souvent destinée à plusieurs acteurs en IE dans des contextes différents : veille, RI, aide à la prise de décision, etc. Dès lors, elle nécessite des adaptations au profit de son usage en adéquation avec les interprétations possibles. Il est question de distinguer les : −conventions adoptées, −langues cible et source, −listes de mnémonique, −valeurs d'un attribut, −indicateurs d'un thème donné, −conventions symboliques, −tableaux de correspondance, etc. ;

- **la dimension identification :** une identification des attributs et des valeurs dans l'objet de l'annotation est nécessaire pour le rendre opérationnel (ou exploitable). A titre d'exemple en phase de RI, l'acteur en IE a besoin de : *(i)* spécifier les termes de sa requête comme pour les éléments attributs et valeurs dans un document, annotation, etc. ; *(ii)* valoriser les résultats de sa RI avec des éléments à valeurs ajoutées dans les annotations : relier les éléments attributs et valeurs entre la requête, les documents de réponse et leurs annotations ;

- **la dimension traduction :** une annotation au niveau de sa structuration intègre des propriétés sur son rôle dans la communication entre l'annotateur (producteur de l'annotation) et les prospecteurs (ie. ceux qui exploitent les annotations). Ces derniers, dans le contexte d'IE, sont des agents humains (veilleur, analyste ou décideur) qui s'outillent dans leur exploitation par des agents logiciels (ie. outils informatiques) pour traduire leurs traitements en objets de décision.

A l'issu de ces réflexions, nous avons intégré ces différentes dimensions dans l'outil annotation pour la RI. A la base de l'expérimentation, nous avons travaillé sur un corpus de test constitué de documents primaires (ie. sources informationnelles), de documents secondaires (ie. notices bibliographiques associées aux sources). Les valeurs ajoutées sont couvertes par l'activité de l'annotateur en déployant la fonction sémantique pour constituer le document tertiaire (ie. annotation associée à la source, à la notice ou à une autre annotation). Des résultats issus de l'expérimentation sur corpus (ie. images botaniques (Yosase, 2007) et images du patrimoine tunisien (Khemiri, 2007) dans le cadre de projets de master en IST-IE à l'université de Nancy2) ont validé les principes de la modélisation.

Dans la construction de l'annotation, on distingue les annotations implicites (ie. des attributs et des valeurs non reliés) et celles explicites (ie. chaque attribut est relié à une ou plusieurs valeurs) à travers les actions de l'annotateur. Pour une annotation implicite, nous avons associé l'objet implicite, *idem*. une annotation explicite est associée à l'objet explicite : le but est d'entreprendre des traitements et des analyses adéquats par type d'objet (Salton et alii., 1994). En phase d'intégration d'un objet, l'annotateur (ie. acteur en IE) fait appel à la fonction sémantique qui s'applique sur un contenu donné (ie. *f°semantique(contenu)*) pour l'annoter. Cette dernière réactive des règles associatives pour la détermination des éléments attributs (A) et/ou valeurs (V) dans l'objet. Finalement, ces règles associatives permettent de statuer si l'objet est implicite ou explicite.

– Pour un objet explicite (*O_explicite*) :

$A \rightarrow <attribut>$ ;
$V \rightarrow <valeur>^{+}$ ;
$O\_explicite \equiv Action\ (f°semantique(contenu)) \rightarrow (A, V)^{+}$ ;
$+ : (1, n)$ ;

– Pour un objet implicite (*O_implicite*) :

$A \rightarrow <attribut>$ ;
$V \rightarrow <valeur>^{+}$ ;
$O\_implicite \equiv Action\ (f°semantique(contenu)) \rightarrow (A*, V*)^{+}$ ;
$* : (0, n)$ ; $\exists\ !\ (A,V) \neq \varnothing$ ;

– Pour l'intégration d'un objet de l'annotation (*O_annotation*) :

$O\_annotation \rightarrow \{O\_implicite*, O\_explicite*\}$ ;

Condition : $\exists\ !\ (O\_implicite, O\_explicite) \neq \varnothing$ ;
*ie. au moins un des deux objets est non vide.*

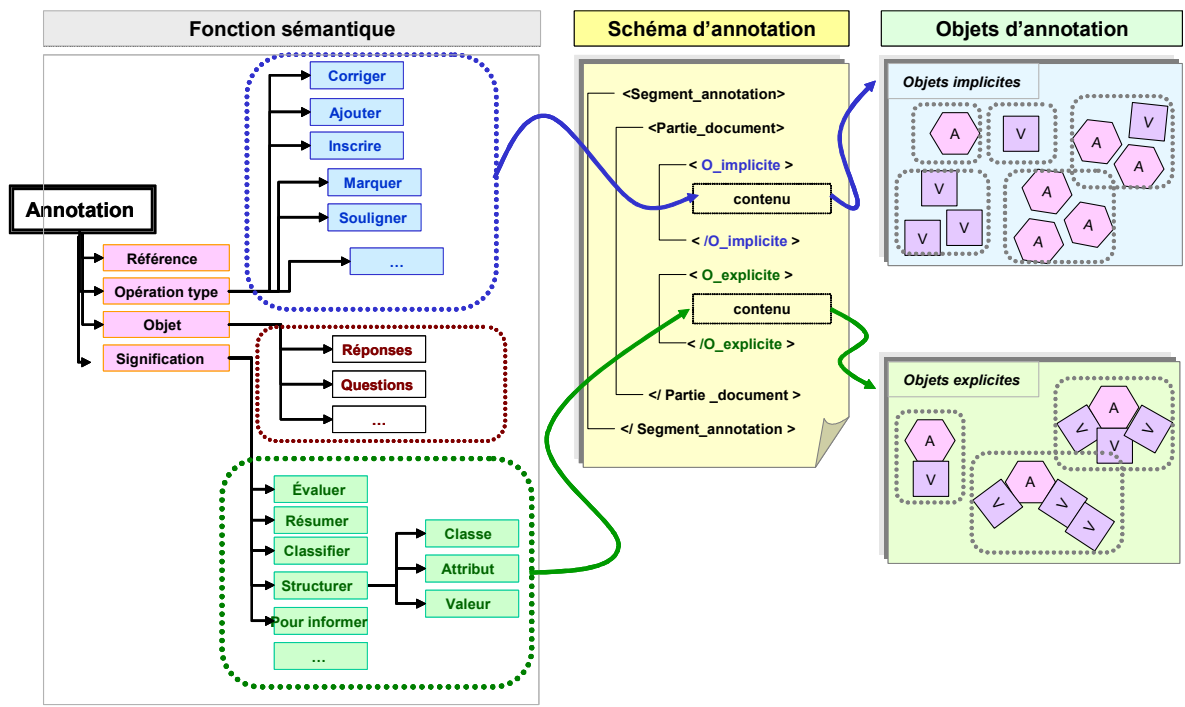

***Figure 2. :*** *Fonction sémantique, schéma et objet de l'annotation.*

Dans la modélisation, nous avons défini des classes sémantiques pour déterminer des caractéristiques à attribuer à l'annotation. Chaque classe regroupe (*cf.* Fig. 2.) les propriétés suivantes :

– **Contexte d'annotation :** la création d'une nouvelle annotation ou le suivi d'une ancienne permet de définir s'il s'agit d'une requête, d'une RI, d'une interprétation, d'une proposition (objet de décision, procédure, reformulation, etc.), etc. ;

– **Document à annoter :** des spécifications portant sur le document à annoter, sa typologie, s'il s'agit d'une source/ notice/ annotation, etc. ;

- **Annotateur :** la description du profil de l'annotateur. Le profil peut être déterminé explicitement dans la base des profils ou implicitement par ses activités de RI ;
- **Annotation :** par cette propriété que l'objet est défini (ie. objet implicite ou explicite) avec des spécifications portant sur la fonction sémantique utilisée, l'opération type, l'objectif, la signification et le contenu (ie. attributs et valeurs) de l'annotation ;

Après la présentation des concepts clés sur l'objet de l'annotation, la principale difficulté réside dans la *détermination "sémantique"* des objets implicites : ce qui nécessite d'impliquer un mode de raisonnement sur les éléments de son contenu $(A^*, V^*)^+$.

Le mode de raisonnement va nous permettre de développer un ensemble de mécanismes logiques sur les attributs ou les valeurs. Il s'agit de confronter les concepts de l'objet implicite à des connaissances préalables $(A, V)^+$ (ie. dans le SRI) afin d'obtenir des associations possibles avec les éléments manquants A ou V. Notre choix s'est orienté vers le raisonnement par inférence pour remédier au problème posé.

## 3.2. Raisonnement par inférence sur l'objet « annotation »

Dans sa définition classique, l'inférence est une opération logique portant sur des propositions tenues pour vraies (ie. les prémisses) et concluant à la vérité d'une nouvelle proposition en vertu de sa liaison avec les prémisses.

Dans cette considération, nos attentes en impliquant un raisonnement par inférence est de pouvoir développer un ensemble de déductions logiques pour identifier l'un des éléments manquants (ie. A ou V) dans l'objet implicite. Ce dernier est confronté à des connaissances préalables (ie. objets explicites) dans la base des annotations pour le rendre "explicite" et continuer les traitements sur les annotations :

- selon la syntaxe Prolog (Colmerauer, 2007), une formalisation de l'annotation, en logique des prédicats du $1^{er}$ ordre, est présentée comme suit :

```
DOMAINS
        (...)
        A= attribut*
        V= valeur*
PREDICATES
        Annotation(attribut, V, C)
        Element_de_Attribut(attribut, A)
        Element_de_Valeur(V, V)
CLAUSES
Annotation(X, Y, C) :- Element_de_Attribut(X, LA) ∧ Element_de_Valeur(Y, LV) ∧ Document(C).

%--
avec:
Annotation(X, Y, Z) :        c'est une  annotation sur le contenu/ document Z qui a comme attribut X
                             de la liste LA et comme valeurs d'attribut la sous-liste Y de LV ;
Exemple :        annoter les mots M₁ à Mₙ comme des concepts-clés dans le document Z,
                 revient à définir dans l'annotation que X est l'attribut "concept-clé" et Y= [M₁,..., Mₙ])
                 les valeurs de l'attribut ;
∃! (X , Y) ≠∅ ; (ie.  au moins un élément de (A,V) est non vide) ;
--%
```

Or dans le schéma d'annotation (*cf.* Fig. 2.), nous avons distingué les objets explicite et implicite. Pour un objet explicite, le domaine du raisonnement ne peut contenir que des faits ou des règles qui infèrent des faits. Il s'agit des associations $(A, V)^+$ entre un attribut et des valeurs, où les attributs et valeurs ont été explicités par l'annotateur.

– selon la syntaxe Prolog, une formalisation de l'objet explicite est présentée par des faits ou des règles comme suit :

```
CLAUSES
%-- Faits : annotations explicites --%
Annotation("souligner", ["stratégie", "développement"],"note_91007").
Annotation("mots clés", ["ATN", "formalisme", "analyse"], "note_71007").
Annotation("ordonner",[    ["pauvre",0],["faible",1],    ["moyen",2],    ["riche",3],    ["pertinent",4]    ],
"note_56007").
etc.

%-- Faits : attributs par génération --%
Generer_attribut([X|LA], Z) :-  Annotation(X,Y,Z) ∧ Generer_attribut(LA, Z).
Generer_attribut([],Z).

%-- Faits : listes de valeurs par génération --%
Generer_valeurs([Y|LV], Z) :-  Annotation(X,Y,Z) ∧ Generer_valeurs(LV, Z).
Generer_valeurs([], Z).
```

En contrepartie pour un objet implicite, le domaine du raisonnement n'est pas défini qu'un élément de $(A^*, V^*)^+$ est manquant. Il est question de faire appel au mécanisme d'inférence pour la détermination de l'élément manquant dans l'objet implicite exploré.

– Pour résoudre ce problème, on peut définir des règles de détermination sur l'un des éléments (A,V) manquants. Selon la syntaxe Prolog, des règles sont définies pour retrouver le paramètre manquant comme suit :

```
CLAUSES
%-- Paramètre manquant : attribut dans O.implicite --%
Expliciter_Annotation(X, Z) :- Element_de_Attribut (X, LA) ∧ Annotation(X, V, Z).

%-- Paramètre manquant : valeurs dans O.implicite --%
Expliciter_Annotation(X, Z) :- Element_de_Valeur (X, LV) ∧ Annotation(A, X, Z).
```

Dans l'environnement actuel du processus d'annotation, la recherche d'information dans les contenus annotés opère par analyse simple ou avec contraintes et permet de formuler : *(i) une recherche classique* en spécifiant les deux paramètres (A,V), *(ii) une recherche avec contraintes* en indiquant seulement un des paramètres de (A,V) et c'est ainsi que le raisonnement par inférence est déclenchée. Nous développons, sur des exemples, les deux modalités de RI dans les annotations comme suit :

– **Pour la recherche classique,** il s'agit de combiner des critères $(A, V)^+$ par des opérateurs booléens. L'expression de recherche : *("auteur", ["Alain Juillet"])* ET *("mots-clés", ["désinformation", "intelligence stratégique", "décision"])*, nous retourne la liste des annotations impliquées ["note_702", "note_008", "note_211"].

– **Pour la recherche avec contraintes,** le mécanisme d'inférence permet de suppléer à l'information manquante dans *(A, V)*. L'expression de recherche *(["désinformation", "protection du patrimoine", "pertinent"])* permet d'inférer :

**(a.)** l'attribut ***"mots-clés"*** pour ***"désinformation"*** et ***"protection du patrimoine"*** *dans la liste de valeurs ["désinformation", "intelligence stratégique", "décision"]* et
**(b.)** l'attribut ***"souligner"*** pour ***"pertinent"*** ou
**(c.)** la valeur ***"pertinent"*** pour l'attribut ***"ordonner"*** qui est associé à la liste de valeurs *[["pauvre",0], ["faible",1], ["moyen",2], ["riche",3], ["**pertinent**",4]]*.

Avec ces possibilités, la recherche d'information est réécrite comme pour une recherche classique (a ET (b OU c)) : *("mots-clés", ["désinformation", "intelligence stratégique", "décision"]) ET (("souligner", ["pertinent"]) OU ("ordonner", [["pauvre",0], ["faible",1], ["moyen",2], ["riche",3], ["pertinent",4]])),* qui nous retourne la liste des annotations impliquées *("note_211").*

Parmi nos objectifs, il est question de mieux manager les connaissances tacites (informelles, non explicitées) entre un groupe d'acteurs en IE qui veulent aller plus loin ensemble à partir de leur savoir mis en commun tout en étant de cultures différentes : l'information pertinente ne s'inscrit pas uniquement dans les ressources primaires ou secondaires et qu'il est essentiel d'intégrer ces acteurs "prospecteurs de l'information" pour relever et annoter des connaissances appropriées au contexte de la prise de décision.

C'est ainsi que le processus d'annotation a été introduit en RI pour parvenir à une modélisation avec un objet fonctionnel pour l'acteur et le système en IE. Le processus intègre les concepts de l'annotation, l'objet (explicite ou implicite) et l'action (annotateur) pour définir la valeur ajoutée sur un contenu et pour aider le décideur à s'en servir dans la résolution d'un problème décisionnel.

## 4. CONCLUSION

Dans le contexte de la société de l'information, la disponibilité des ressources informationnelles est aujourd'hui une question très sensible, car l'information est devenue une source de pouvoir. Pour les organisations, l'information est pondérée aux facteurs de *célérité* (ie. le temps de réponse) et de *pertinence* (ie. une sensibilité variable qui tend à l'appréciation par les acteurs) par rapport aux *besoins informationnels des décideurs*. Le rapprochement entre un problème décisionnel et sa résolution ne peut s'établir sans la synergie des acteurs en IE : leurs sensibilités vis-à-vis de la valeur informationnelle, et des outils pour gérer et fournir l'information et ses valeurs. Cette sensibilité est maîtrisée par l'intégration du processus d'annotation : les *actions* de l'annotateur à la construction des *objets* de l'annotation qui tend vers l'induction de la valeur ajoutée sur les contenus informationnels. En conséquence, l'annotation ne peut qu'introduire un niveau supplémentaire dans l'exploitation de l'information, qui n'est pas de l'auteur mais des acteurs « butineurs » de cette information dans leur contexte (*cf.* Fig. 3.).

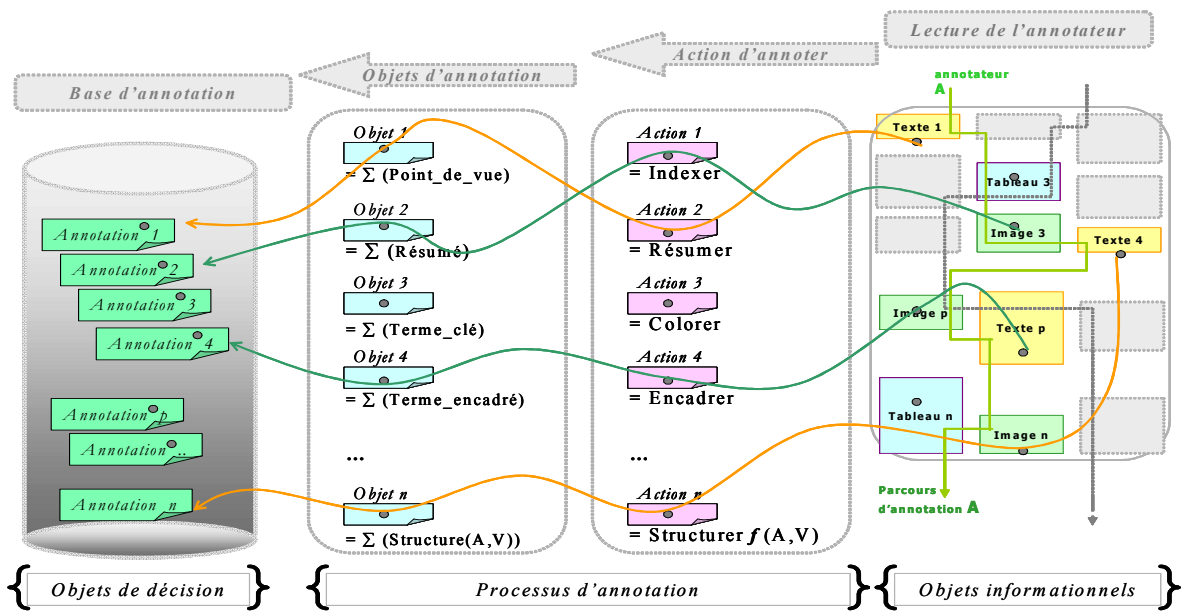

***Figure 3.** : Processus d'annotation pour l'aide à la prise de décision.*

Dans cette perspective, une annotation sur un document peut avoir une visibilité et une diffusion plus large que le spectre réservé par l'auteur du document, car il convient de rappeler que le concept de "la valeur" est « *une notion relativement subjective, circonstancielle, dépendant d'une intention et d'une compétence* » (Portnoff, 2005). Ainsi, l'annotation, sous toutes ses formes (structurée ou non, implicite ou non et déterministe ou non), permet dans une optique de capitalisation de connaissances d'expliciter des propositions à sémantique forte sur une ressource informationnelle, un contenu documentaire, une information, une connaissance, etc.

Ainsi, on peut considérer que l'objet d'annotation est une résultante de multiples actions et interprétations des annotateurs, qui ne sont pas nécessairement conformistes aux idées transmises par l'auteur de l'information, car le contexte de ce dernier n'est pas celui des acteurs et du décideur vis-à-vis d'un problème décisionnel en IE.

Dans l'optique de coordonner les actions des acteurs en IE sur un problème décisionnel, les caractéristiques intrinsèques d'une annotation sont très proches de celles du « *document pour l'action* » (Zacklad, 2004), à savoir :

- un objet annotation est le support, la visée ou le résultat d'action dans une approche de prospective ou exploratoire de l'information ;

- une certaine pérennité associée aux engagements de l'annotateur à l'égard d'un contenu véhiculé : idées, connaissances, savoir-faire, confrontation, évaluation, etc. ;

- un statut d'inachèvement prolongé : comme sur un document, une annotation peut aussi être annotée, comme elle peut être modifiée pour s'enrichir dans le temps par d'autres annotateurs;

- plusieurs annotateurs aux rôles et à l'engagement éventuellement très différents et évolutifs : profils intellectuel, psychologique et socio-professionnel des acteurs en IE (Bouaka, 2004) ;

- le contenu d'une annotation est susceptible de multiples interprétations : si l'annotateur ne structure pas son annotation, elle restera sujette à des interprétations et à des déductions possibles selon la logique de raisonnement (ie. mécanisme d'inférence, principe d'induction, etc.).

En perspective, un aspect fondamental, qui peut être pris en considération dans le processus d'annotation, consiste à transmettre au décideur des informations décisionnelles "anticipées" (ie. les objets de décision par l'annotation) :

*(i)* des informations qui répondent à un besoin informationnel (ie. fournies par le SRI),
*(ii)* des annotations associées à ces informations (ie. fournies par les acteurs en IE) qui répondent à l'explicitation d'un problème décisionnel.

Dans cette orientation, l'annotateur a un rôle capital, essentiellement quand le processus intègre les actions du décideur par ses explicitations, ses interprétations et ses motivations.

Finalement, certaines questions sur le concept d'annotation restent ouvertes et méritent encore des efforts d'étude pour la continuité des propositions : apporter une solution au décideur ne peut contourner une culture « partagée » entres les acteurs en IE et que l'information n'a pas de valeur en soi.

# RÉFÉRENCES